# Relating Changes in Cometary Rotation to Activity: Current Status and Applications to Comet C/2012 S1 (ISON)

Nalin H. Samarasinha[1] and Béatrice E. A. Mueller[1]

[1] Planetary Science Institute
1700 E Fort Lowell Road, Suite #106, Tucson, AZ 85719, USA.

Running Head: Cometary Rotation, its Changes, and Activity

# ABSTRACT

We introduce a parameter, *X*, to predict the changes in the rotational period of a comet in terms of the rotational period itself, the nuclear radius, and the orbital characteristics. We show that *X* should be a constant if the bulk densities and shapes of nuclei are nearly identical and the activity patterns are similar for all comets. For four nuclei for which rotational changes are well documented, despite the nearly factor 30 variation observed among the effective active fractions of these comets, *X* is constant to within a factor two. We present an analysis for the sungrazing comet C/2012 S1 (ISON) to explore what rotational changes it could undergo during the upcoming perihelion passage where its perihelion distance will be ~2.7 solar radii. When close to the sun, barring a catastrophic disruption of the nucleus, the activity of ISON will be sufficiently strong to put the nucleus into a non-principal-axis rotational state and observable changes to the rotational period should also occur. Additional causes for rotational state changes near perihelion for ISON are tidal torques caused by the sun and the significant mass loss due to a number of mechanisms resulting in alterations to the moments of inertia of the nucleus.

*Key words:* comets: general, comets: individual (C/2012 S1 (ISON))

# 1. INTRODUCTION

The determination of accurate rotational parameters for cometary nuclei, let alone their changes over time, is a challenging task. When comets are inactive or barely active, the scattered light from the nucleus dominates the flux rather than the coma and the corresponding rotational lightcurves yield rotational periods. By combining lightcurves taken sufficiently apart in time (e.g., of the order of a month) but having approximately the same Earth-comet-sun geometry, one can determine rotational periods with high precision. However, such rotational lightcurves of the nucleus can be obtained only when the comet is inactive or barely active at large heliocentric distances. At such distances, the nucleus should be sufficiently large to produce rotational lightcurves with good signal-to-noise. As comets get closer to the sun, they become active and the flux is dominated by the coma. Then, obtaining a rotational lightcurve becomes infeasible unless the lightcurve is modulated by activity (e.g., 1P/Halley; Millis & Schleicher 1986) where activity here represents nuclear outgassing. If a bright comet is relatively close to the observer and shows identifiable coma features, repeatability of those features can be used to derive rotational periods. These observational challenges limit the number of comets with known rotational parameters (e.g., Samarasinha et al. 2004, and references therein).

However, during the last decade, partly due to cometary flyby missions and associated global observational campaigns, we now have definitive confirmation for changes in rotational periods over orbital timescales or shorter (see Section 2 for details). These changes occur primarily due to reaction torques from volatile outgassing from the nucleus (changes to the moments of inertia due to mass loss could be a secondary contributor at least for the relatively small comet 103P/Hartley 2, and also comet C/2012 S1 (ISON)).

We introduce a parameter $X$ relating the changes in rotational period to activity. We show that $X$ is nearly constant for the four comets for which we have observational confirmation of rotational changes despite their widely different activity levels. An analysis of how activity may influence the sungrazing comet C/2012 S1 (ISON) will be presented and what kind of rotational changes that one can expect will also be explored.

# 2. OBSERVATIONAL EVIDENCE FOR CHANGES IN THE ROTATIONAL PERIODS OF COMETS

There are four short-period comets (three of which are Jupiter-family and two of which had flyby missions) for which we have confirmed determinations of changes in their rotational states. They are comets 2P/Encke, 9P/Tempel 1, 10P/Tempel 2, and 103P/Hartley 2. Observationally, the easiest monitor of the rotational changes is the rotational period. Sometimes, changes to the direction of the total rotational angular momentum vector (hereafter TRAM vector) can be determined (e.g., 19P/Borrelly; Schleicher et al. 2003,

Farnham & Cochran 2002); however, in general, determining changes in the TRAM vector is more challenging than determining the changes in the rotational periods.

*Comet 2P/Encke:* 2P/Encke has an orbital period of 3.3 years and is therefore one of the most extensively observed comets. It is an enigmatic object in many ways. Foremost is the unusual behavior of its brightness when 2P/Encke is apparently not active. It brightens when 2P/Encke is near aphelion (e.g., Belton et al. 2005, Meech & Svoren 2004). Also, 2P/Encke is extremely dust-poor at visible wavelengths (e.g., A'Hearn et al. 1995), suggesting that the number of micron-sized grains is depleted when compared to other comets. However, the presence of a dust trail (e.g., Reach et al. 2000) indicates that 2P/Encke releases a prodigious amount of mm-sized and larger grains. The rotational period of 2P/Encke is ~11 hours (Harmon & Nolan 2005, Fernández et al. 2005) and it is spinning down by ~4 minutes per orbit assuming a constant rate of change. These changes have been observed in more than one apparition (Mueller et al. 2008). The effective nuclear radius is ~2.4 km (Fernández et al. 2000).

*Comet 9P/Tempel 1:* As 9P/Tempel 1 was twice a NASA mission target (for Deep Impact and Stardust-NExT), it had supporting ground-based observing campaigns (e.g., Meech et al. 2011a, 2005 making 9P/Tempel 1 a well-observed comet. The rotational period of the comet is ~41 hours (Chesley et al. 2013, Belton et al. 2011). Chesley et al. (2013) deduced that 9P/Tempel 1 spun up by either 12 minutes or 17 minutes during the 2000 perihelion passage and by 13 minutes during the 2005 perihelion passage, building on the work by Belton et al. (2011). The effective nuclear radius is ~2.8 km (Thomas et al. 2013a).

*Comet 10P/Tempel 2:* Due to its relatively large, yet elongated nucleus (e.g., Lamy et al. 2004, and references therein), 10P/Tempel 2 is well suited for obtaining rotational lightcurves of the nucleus. The rotational period of the comet is ~9 hours (Knight et al. 2012, 2011, Mueller & Ferrin 1996, Sekanina 1991). It is spinning down at a rate of 16 seconds per perihelion passage assuming a constant rate of change (Knight et al. 2012, 2011). The effective nuclear radius is ~6.0 km (Lamy et al. 2009).

*Comet 103P/Hartley 2:* 103P/Hartley 2 had a close approach to Earth (0.12 AU) and was the cometary target of NASA's EPOXI mission with an encounter on November 4, 2010. The comet has a very small (effective radius ~0.58 km), irregular, but highly active nucleus (Thomas et al. 2013b, A'Hearn et al. 2011). The comet's observed rotational period[1] changed from 16.6 hours in August to 17.1 hours in early September to 17.6 hours in early October to 18.1 hours in late October to 18.4 hours in early November, and to 18.8 hours in mid-November (Belton et al. 2013, Samarasinha et al. 2011, 2010, Meech et al. 2011b, A'Hearn et al. 2011, Knight & Schleicher 2011, Drahus et al. 2011), a change of almost 2 hours in 3 months! This comet also showed evidence for rotational excitation.

---

[1] Comet 103P/Hartley 2 is in a non-principal-axis rotational state. However, as this is observationally manifesting as a minor deviation from a principal-axis rotational state, the rotational period quoted is the component corresponding to the motion of the long axis around the TRAM vector (i.e., the dominant component period).

Rotational changes were also reported for non-periodic comets C/2001 K5 (LINEAR) (Drahus & Waniak 2006) and C/1990 K1 (Levy) (Feldman et al. 1992, Schleicher et al. 1991). However, as they were only observed during a limited time during their active phase, they are not considered in our analysis.

Table 1 summarizes the rotational periods, $P$, for the above short-period comets and the respective changes in the rotational periods over an orbit, $\Delta P$. Note that $\Delta P$ for 9P/Tempel 1 is not constant and 103P/Hartley 2 is in an excited rotational state.

Table 1: Short-period Comets with Confirmed Rotational Changes

| comet | P [hours] | ΔP [minutes] |
|---|---|---|
| 2P/Encke | 11 | 4 |
| 9P/Tempel 1 | 41 | -14 |
| 10P/Tempel 2 | 9 | 0.27 |
| 103P/Hartley 2 | 18 | 120 |

# 3. $X$: A PARAMETER TO INTERPRET CHANGES IN THE ROTATIONAL PERIODS

To interpret changes in the rotational periods due to outgassing torques, we develop a simple theoretical framework starting with a few basic assumptions and then we explore the robustness of those assumptions and their implications.

The change in the angular velocity of the nucleus due to reaction torques is given by

$$I\dot{\omega} = N \qquad (1)$$

where $I$ is the moment of inertia of the principal axis around which the nucleus rotates, $\dot{\omega}$ is the rate of change of the angular velocity, and $N$ is the net component torque due to outgassing responsible for changing the rotational rate. $N$ and $I$ can be expressed as follows.

$$N = c\,|\mathbf{R_n} \times Q\mathbf{V}| \qquad (2)$$

where $c$ is a constant, $\mathbf{R_n}$ is the nucleus radius vector, $\mathbf{V}$ is the gas outflow velocity at the nucleus[2], and $Q$ is the gas production rate (measured in molecules per unit time) at the nucleus. Here, integration over the entire nucleus is implied.

---

[2] This is the gas outflow velocity *at the nucleus* and not in the coma after gas acceleration due to photochemical heating (e.g., Combi et al. 1999) and may correspond to the sublimation temperature of the ice and therefore exhibits only a weak dependence on the heliocentric distance.

$$I = \lambda \rho R_n^5 \tag{3}$$

where $\lambda$ is a shape dependent coefficient and $\rho$ is the bulk density of the nucleus.

If all nuclei have similar bulk densities and the torques exerted on all nuclei by the escaping gases are in the same proportion to the production rates of gases at the respective nuclei, then

$$\left|\dot{\omega}\right| = C \frac{Q}{R_n^4} \tag{4}$$

where $C$ is a constant.

Since

$$\omega = \frac{2\pi}{P} \tag{5}$$

where $P$ is the rotational period, then

$$\left|\dot{\omega}\right| = 2\pi \frac{\left|\dot{P}\right|}{P^2} \tag{6}$$

where $\dot{P}$ is the time derivative of $P$. From equations (4) and (6), we have

$$\left|\dot{P}\right| = k \frac{P^2 Q}{R_n^4} \tag{7}$$

where $k$ is a constant.

However, instead of $\dot{P}$, observations yield the change in period, $\Delta P$, over a time interval when the nucleus is clearly active during its orbit, say <3 AU. Furthermore, the only strongly time-dependent quantity in equation (7) is $Q$. Therefore, assuming $\dot{P}$ has the same sign over the entire active orbit,

$$\left|\Delta P\right| = K \frac{P^2 \int_{active\ phase} Q(t)\, dt}{R_n^4} \tag{8}$$

where $K$ is a constant.

If all the nuclear shapes are nearly identical and the surface distribution of activity is roughly the same, then

$$\int_{active\ phase} Q(t)\, dt = h f R_n^2 \zeta \tag{9}$$

where $h$ is a constant, $f$ is the effective active fraction, and

$$\zeta = \int_{active\ phase} Z(t)\,dt \tag{10}$$

with $Z(t)$ being the water production rate at the nucleus per unit surface area at zero solar zenith angle and depends on the heliocentric distance of the comet, $r_h$.

The effective active fraction, $f$ (as a percentage of the total surface area of the nucleus) is the sum of (a) the "active fraction" based on the amount of water directly sublimating form the nucleus, $f_{dw}$, and (b) the 'water-equivalent active fraction" corresponding to other gases directly sublimating from the nucleus, evaluated as if the corresponding reaction force (and torque) is due to an equivalent water sublimation directly from the nucleus, $f_{og}$.

From equations (8) and (9), we have

$$|\Delta P| = H \frac{P^2 f \zeta}{R_n^2} \tag{11}$$

where $H$ is a constant.

If $f$ is the same for all comets, then

$$X = \frac{|\Delta P| R_n^2}{P^2 \zeta} = \text{constant}. \tag{12}$$

Table 2 lists $|\Delta P|$, $R_n$, $P$, $\zeta$, $X$ (normalized to the values for 2P/Encke) and the effective active fractions, $f$, for the four short-period comets discussed above. After examination of all the assumptions made, we conclude that the least justifiable assumption is that all comets have the same $f$. However, we find $X$ is indeed constant, varying only by approximately factor two despite the nearly factor 30 variation in the effective active fractions!

Table 2: Parameters for Relating Changes in Rotational Periods to Nuclear Activity for Short-Period Comets

| comet | $|\Delta P|/|\Delta P_{Encke}|$ | $R_n/R_{nEncke}$ | $P/P_{Encke}$ | $\zeta/\zeta_{Encke}$ [a] | $X/X_{Encke}$ | $f$ [%] |
|---|---|---|---|---|---|---|
| 2P/Encke | 1 | 1 | 1 | 1 | 1 | 2 [b] |
| 9P/Tempel 1 | 3.50 | 1.17 | 3.72 | 0.32 | 1.1 | 5 [b] |
| 10P/Tempel 2 | 0.068 | 2.50 | 0.82 | 0.34 | 1.9 | 0.6 [b] |
| 103P/Hartley 2 | 30.0 | 0.24 | 1.63 | 0.42 | 1.5 | 20 [c] |

**Notes.**
[a] $Z(t)$ for a given $r_h$ is based on the thermo-physical model of Gutiérrez et al. (2001).
[b] These values are based on the equivalent median surface area needed to produce the observed water production rates and was derived in a consistent and uniform manner (David Schleicher, personal communication). All observed water is assumed to be due to direct sublimation from the nucleus and since water is the dominant volatile, $f_{og} \ll f_{dw}$ (i.e., $f \approx f_{dw}$).
[c] The observed water production for 103P/Hartley 2 from the 2010 apparition corresponds to 50% active fraction (David Schleicher, personal communication) and the observed water

production decreased over the last few apparitions (e.g., Knight and Schleicher 2013, Combi et al. 2011). However, only a small fraction of water is directly sublimating from the nucleus (Fougere et al. 2013, Belton 2012, Meech et al. 2011b) and ~4% representing the geometrical mean of the range cited in the literature is adopted for $f_{dw}$. The water-equivalent active fraction for other gasses, $f_{og}$, corresponding to $CO_2$ is ~16%. Therefore, the effective active fraction, $f$, for 103P/Hartley 2 is ~20%.

We encourage observations of additional comets to determine their rotational changes and to assess the respective values of $X$. *If indeed X is nearly constant, then the amount of change in the rotational period per orbit can be predicted, irrespective of the effective active fraction!*

*X is a measure of the change in the total rotational angular momentum of the nucleus per unit nuclear mass over the active orbit* normalized to the orbital mass loss of volatiles per unit surface area due to insolation. This normalization is meant to remove the differences in insolation among different cometary orbits. The constancy of *X* implies that for two cometary nuclei of the same size, rotational period, bulk density, shape, and orbital characteristics but with different effective active fractions, the changes to their rotational periods are nearly the same. Therefore, the net torques due to outgassing should be nearly the same irrespective of the effective active fractions. We suggest especially for highly active nuclei, either there are significant cancellations of outgassing torques corresponding to individual surface facets during each diurnal cycle and/or there should be both decreases and increases in the rotational period during an orbital cycle thus negating the full effect of the rotational changes due to outgassing torques.

Therefore, nuclei with high effective active fractions are not necessarily more efficient at changing their rotational periods (provided they are nearly identical in all other aspects). If *X* is nearly constant, $|\Delta P|$ for 1P/Halley is ~20 minutes and is smaller than the accuracy of the current determinations of the component rotational periods[3].

# 4. ANTICIPATED ROTATIONAL CHANGES IN COMET C/2012 S1 (ISON)

Comet C/2012 S1 (ISON) (hereafter ISON) was discovered at $r_h$>6AU and its perihelion distance will be 2.7 solar radii (0.0125 AU) on November 28, 2013. Its orbital eccentricity is 1.000004 and the inclination is 61°.88 suggesting a dynamically new comet of possible Oort Cloud origin. This is the first sungrazing comet to be monitored over a large range of $r_h$ inbound to study its activity and other evolutionary effects as a function of $r_h$. The HST images taken in April 2013 indicate that the effective nuclear radius of ISON is <2 km (Li et al. 2013). The rotational period, *P*, of the comet has not yet been determined.

---

[3] 1P/Halley is in a non-principal-axis rotation.

Due to ISON's small perihelion distance, barring a catastrophic disruption of the nucleus, it should experience more extreme outgassing torques than a non-sungrazer. We present an analysis based on the parameter $X$ as well as numerical simulations to explore the likely rotational changes that ISON might undergo due to outgassing torques.

Assuming $X$ is the same for ISON as that for other comets discussed earlier and since $\zeta/\zeta_{Encke} \approx 5.8$ based on its orbit, we estimate the expected change in its rotational period, $\Delta P$.

Since $X/X_{Encke} \approx 1$, from equation (12) we have

$$\Delta P \approx \Delta P_{Encke} \frac{(\zeta/\zeta_{Encke})(P/P_{Encke})^2}{(R_n/R_{n\,Encke})^2} \ . \tag{13}$$

Table 3 lists the values predicted for $\Delta P$ for different $R_n$ and $P$. The changes in $\Delta P$ are easily detectable. When $R_n < 1$ km, the expected $\Delta P$ could be of the order of $P$ and rotational splitting is likely to be a leading cause for the disruption of small (sub-km size) sungrazers.

Table 3: Anticipated values of $\Delta P$ for ISON

|  | $R_n$=1 km | $R_n$=2 km |
|---|---|---|
| $P$=6 hr | $\Delta P \approx 0.7$ hr | $\Delta P \approx 0.2$ hr |
| $P$=12 hr | $\Delta P \approx 2.7$ hr | $\Delta P \approx 0.7$ hr |
| $P$=24 hr | $\Delta P \approx 11$ hr | $\Delta P \approx 2.7$ hr |

As another approach, we carried out numerical simulations for a range of nuclear radii and initial rotational periods to assess the changes that one may observe in the rotational state and when they may occur in the orbit.

The degree of rotational excitation, $\varepsilon$, defined as (Gutiérrez et al. 2002, Samarasinha 2007)

$$\varepsilon = \frac{I_s - (M^2/2E)}{I_s - I_l} \tag{14}$$

quantifies the extent of rotational excitation. $I_s$ and $I_l$ are the respective moments of inertia components corresponding to the short and long principal axes, $M$ is the total rotational angular momentum, and $E$ is the rotational kinetic energy. $\varepsilon=0$ corresponds to the principal-axis rotation around the short axis whereas $\varepsilon=1$ represents the principal-axis rotation around the long axis. $\varepsilon>0$ corresponds to excited rotational states. The period corresponding to the motion of the long axis around the TRAM vector, $P_\phi$ (e.g., Samarasinha and A'Hearn 1991), provides a proxy for the changes in the magnitude of $M$. Figure 1 shows the temporal evolution of $\varepsilon$ and $P_\phi$ for nuclei of varying radii and initial rotational periods in the orbit of ISON. The simulations used a shape model for 9P/Tempel 1 (but scaled based on the effective radii adopted). We expect the qualitative results and the broader conclusions based on the evolution of the rotational motion to be valid for other shapes too. However, the detailed quantitative results will clearly differ.

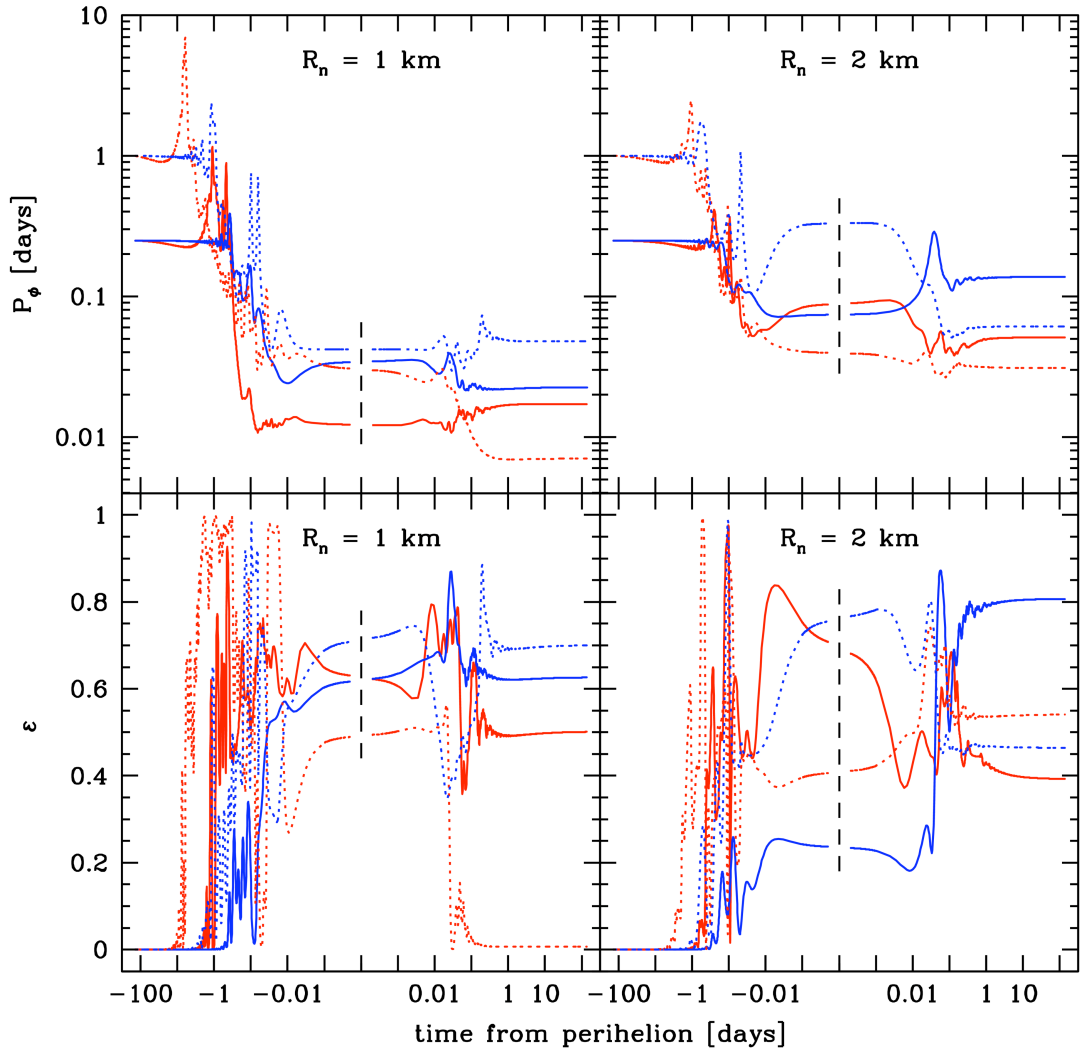

**Figure 1.** Results from numerical simulations showing the temporal evolution of $P_\phi$ (top) and $\varepsilon$ (bottom) for a nucleus in ISON's orbit. Simulations show results for $R_n$ of 1 km (left) and 2 km (right) and initial rotational periods of 6 hours (solid lines) and 24 hours (dotted lines). These simulations were carried out for effective active fractions of 100% (red) and 25% (blue). The water outgassing is occurring from each facet of the entire nucleus based on the insolation that each facet receives normalized to the corresponding effective active fraction. The black dashed lines depict perihelion. Simulations assume a constant mass and the same principal moments of inertia for the nucleus independent of time. However, these assumptions do not hold after about a day prior to the perihelion and the detailed evolution thereafter should be interpreted with extreme caution.

The rotational changes to the nucleus are significant for simulations in Figure 1 and the comet will experience easily observable changes in its rotational state in agreement with the conclusions based on Table 3. The nuclei change their rotational states significantly when near perihelion (typically within a few days) and even for a $R_n$=2 km nucleus, the outgassing torques near perihelion are sufficiently strong to put it into a non-principal-axis rotation.

The fractional mass loss of the nucleus due to activity, $f_m$, can be expressed by the ratio of the volatile (and associated dust) mass loss due to outgassing over the active orbital phase divided by the total mass of the comet.

I.e.,
$$f_m \approx \frac{\zeta f_w \pi R_n^2 / \eta}{4\pi R_n^3 \rho / 3} = \frac{3\zeta f_w}{4\rho \eta R_n} \qquad (15)$$

where $f_w$ is the active fraction corresponding to the total water production rate and $\eta$ is the volatile fraction by mass of the nucleus. For simulations shown in Figure 1, $f_m$ can vary from a few percent to tens of percent depending on $R_n$ and $f_w$. The fractional mass loss due to activity is larger for smaller nuclei.

For our simulations, we did not restrict the range for the rotational rate of the nucleus but let it evolve based on the outgassing torques. In reality, a nucleus will experience rotational splitting at high rotational rates due to centrifugal forces (e.g., Samarasinha et al. 2004, and references therein). Therefore, we expect the nucleus to undergo rotational splitting when it rotates faster than a critical limit[4]. The resultant mass losses will put the nucleus into an excited rotational state due to the altered moments of inertia. Thermal disintegration of ISON due to its small $r_h$ when near perihelion is also a possibility; however, at least for asteroids with near-sun perihelia, there is no observational evidence for detectable mass loss due to thermal disintegration of asteroidal surfaces (Jewitt 2013) except for (3200) Phaethon (Jewitt et al. 2013).

The tidal torques induced by the sun and mass loss due to solar tides, especially when near perihelion, could also influence the rotational state of ISON. A detailed analysis of these effects is beyond the scope of this *Letter*. However, tidal effects could be significant enough to alter the rotational state. For example, Scheeres et al. (2000) show that an asteroid can undergo rotational changes due to tidal torques if within a few planetary radii during a close encounter.

## 5. SUMMARY

1. We introduce a parameter, *X*, which relates changes in the rotational period of a comet to its nuclear activity. *X* should vary among comets depending on the effective active fraction of the nucleus. However, for four comets that we have reliable evidence for changes in their rotational periods, *X* is constant within a factor two despite nearly a factor 30 variation among the effective active fractions.

---

[4] This critical value is of the order of a few hours and depends on the tensile strength of the nucleus (e.g., Samarasinha et al. 2004, Weissman et al. 2004).

2. We urge observations of additional cometary nuclei to confirm or refute this result. If indeed $X$ is nearly constant, then changes in the rotational periods of comets can be easily estimated. If $X$ is nearly constant that may mean significant cancellation of outgassing torques for highly active nuclei.

3. We show that large changes to the rotational state of ISON should occur during its perihelion passage. In addition to changes in the rotational period, the nucleus will be excited into a non-principal-axis rotational state. If the nucleus is in a relaxed rotational state during the post-perihelion, constraints can be placed on the damping timescale and the structural parameters of the nucleus (cf. Samarasinha et al. 2004). There will be significant mass loss from the nucleus and the fractional mass loss could be extremely large for small (e.g., sub-km) nuclei. If ISON survives the perihelion passage, its rotational state will be entirely different from the early pre-perihelion rotational state.

N.H.S. and B.E.A.M acknowledge NASA grants NNX09AM03G and NNX12AG56G for supporting this work. We thank P.J. Gutiérrez for providing water production rates based on his thermo-physical model and D. Schleicher for providing the median surface areas. We appreciate comments by M.F. A'Hearn and P. Tricarico on an earlier draft. We thank the anonymous reviewer for the extremely thorough review. This is PSI Contribution Number 610.